\documentclass[a4paper]{VisionStyle}
\usepackage{epsfig}

\begin{document}

\title{Chandra Observations and the Mass Distribution of EMSS1358+6245:\
    Toward Constraints on Properties of Dark Matter}

\author{M.W.\,Bautz\inst{1} \and J.S.\,Arabadjis\inst{1} \and 
G.P.\,Garmire\inst{2} } 

\institute{ 
Center for Space Research, Massachusetts Institute of Technology, Cambridge, MA 02139 USA
\and 
  Department of Astronomy and Astrophysics, The Pennsylvania State University, University Park, PA 16802 USA} 

\maketitle 

\begin{abstract}
Chandra observations of lensing galaxy clusters have 
now provided accurate dark matter profiles for several objects in which
the intracluster medium is likely to be in hydrostatic equilibrium.
We discuss Chandra observations of the mass profile of one such cluster,
EMSS1358+6245. We find no evidence 
for flattening of the mass density profile at radii greater than 
$50$ h$_{50}^{-1}$ kpc . This result, and 
similar findings from Chandra observations of other clusters, appear
to rule out models in of dark matter self-interaction proposed to
explain  the flat cores of low-surface brightness galaxies.

\end{abstract}

\section{Introduction}
\label{mbautz-B3-sec:intro}
X-ray emission from the intracluster medium (ICM) is a powerful diagnostic of 
the structure of galaxy clusters. If the ICM
has reached  hydrostatic equilibrium  within the cluster gravitational 
potential, one can infer the run of thermal 
pressure  with radius from X-ray imaging spectroscopy, and so obtain the 
cluster mass distribution.  The assumption of hydrostatic equilibrium, as well 
as the reliability of the estimated mass enclosed within fiducial radii, can be checked 
by comparison with other mass constraints obtained, for example, from 
gravitational lensing. Despite some famous cases of diasagreement between 
lensing and X-ray mass estimates, it was shown 
by \cite*{mbautz-B3:all98} that  the X-ray data themselves can be used (along with 
accurate optical cluster positions) to select objects
which are likely to be in hydrostatic equilibrium.  Morevover, the superb 
angular resolution of the current
generation of X-ray observatories makes it relatively straightforward to detect
departures from hydrostatic equilibrium (e.g. \cite{mbautz-B3:mar00};\cite{mbautz-B3:vik01};
\cite{mbautz-B3:mac02}) in the ICM.

Our purpose here is to note that this new observational capability, 
which provides not only accurate masses, but also accurate mass distributions 
for clusters, may actually lead to constraints on  the nature of dark matter itself. 
To illustrate this point we discuss Chandra results for the cluster
EMSS1358+6245.  These results are described in detail elsewhere
(\cite{mbautz-B3:jsa01}) so we provide only a summary here.

\section{Observations and Analysis}
\label{mbautz-B3-sec:obs}

EMSS1358+6245, at redshift z=0.33, is a luminous (L$_{X} = 7 \times 10^{44}$ erg 
s$^{-1}$, 0.2 - 4.5 keV; \cite{mbautz-B3:fmb91}) 
cluster in which a cooling flow (\cite{mbautz-B3:mwb97}; \cite{mbautz-B3:all98})
has been reported.  Both strong (\cite{mbautz-B3:fra97}) and weak 
(\cite{mbautz-B3:hoe98})  gravitational lensing have been used to 
constrain the cluster mass distribution.
The mean ASCA temperature (7 keV; \cite{mbautz-B3:all98}) and the ROSAT surface brightness 
profile together imply a mass of about $4 \times 10^{14} M_{\sun}$, consistent
with the lensing studies.  

\begin{figure}
\centerline{\epsfig{file=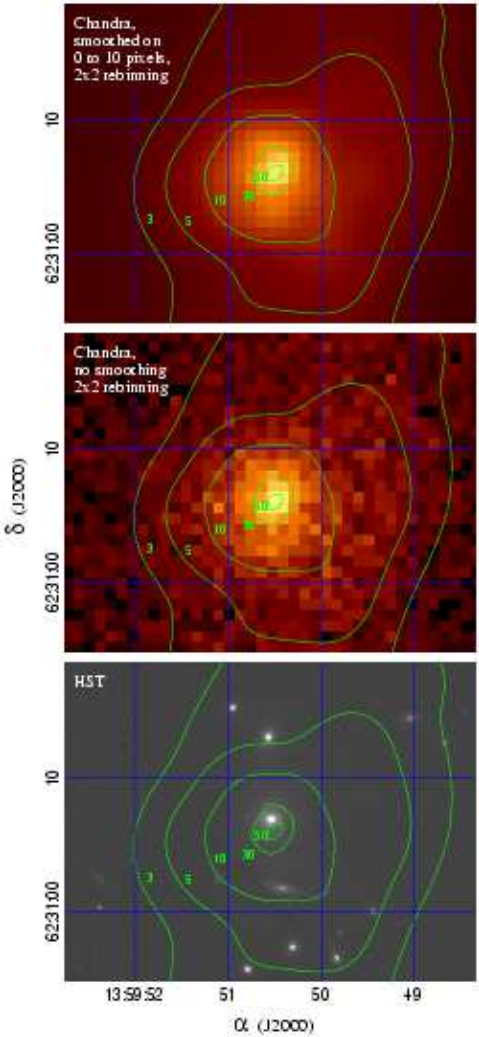,width=7cm}}
\caption{Adaptively smoothed (top) and raw (middle) 0.3-7 keV Chandra images of
and HST image (bottom) of the core of EMSS1358+6245. X-ray photons have been grouped
in $2\times2$ pixel (0.98\,\arcsec$\times$0.98\,\arcsec) bins. X-ray
surface brightness contours derived from the top image are overlaid on all images;
the contours are labeled in units of photons per
0.98\,\arcsec$\times$0.98\,\arcsec pixel.  The X-ray peak coincides
with the center of the brightest cluster galaxy within the estimated
1\,\arcsec astrometric errors in these data.}
\label{mbautz-B3-fig:fig0}
\end{figure}

We summarize here a  55 ks Chandra observation of EMSS1358+6245 with the ACIS/S3
detector performed in 2000 September. Smoothed and raw Chandra images of the cluster
core are shown and compared with an HST optical image in
 Figure~\ref{mbautz-B3-fig:fig0}.
We confirm the presence of a low-temperature, high-X-ray-surface-brightness 
component at the center of the 
cluster ($r < 65$ h$_{50}^{-1}$ kpc). The peak of the X-ray
emission coincides with the location of the brightest cluster galaxy within
the astrometric errors of about 1\arcsec\ (6~h$_{50}^{-1}$ kpc), and the isophotes
are roughly circular.  These characteristics suggest that the ICM is likely to be
in hydrostatic equilibrium in the  cluster gravitational potential
(\cite{mbautz-B3:all98}) so that the X-ray data will yield a reliable cluster
mass profile.

\begin{figure}
\centerline{\epsfig{file=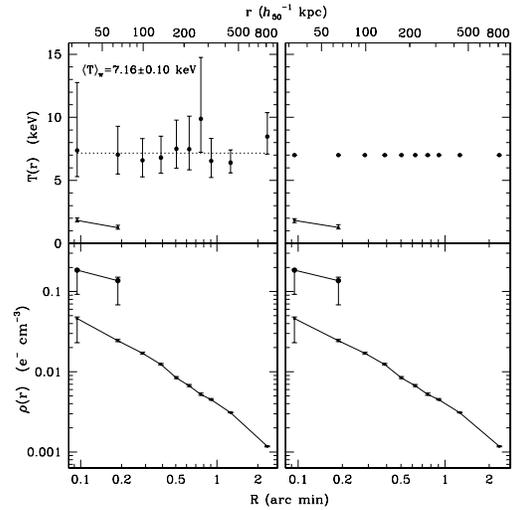,width=7cm}}
\caption{Temperature and electron density of EMSS1358+6245 inferred from
a deprojection analysis of Chandra data. In each of the two central bins 
a two-temperature model was required by the data.}
\label{mbautz-B3-fig:fig1}
\end{figure}

We have fit spatially resolved 
Chandra spectra from 10 concentric annuli simultaneously for the electron 
density and temperature of the ICM in each of 10 corresponding concentric, 
spherical shells.  We used the MEKAL model in XSPEC, fixing the metal abundance 
at 0.3 solar, and  fit for a single Galactic column common to all annuli.
We obtain an acceptable fit to the Chandra data with a two-temperature model
in each of the two central bins (which have
$r < 11\,$\arcsec$\,=  65$ h$_{50}^{-1}$ kpc); a single-temperature model fits
well at larger radii.  We have assumed pressure equilibrium between the hot and
cool components in order to determine their densities.  Results are shown in
 Figure~\ref{mbautz-B3-fig:fig1}.
As expected from the ASCA and ROSAT data, a
single-temperature component does not fit the data well in the cluster core. 

\begin{figure*}[!ht] 
%\centerline{\epsfig{file=ms1358_Mr_floatT_nobox.ps,width=15cm}}
\centerline{\epsfig{file=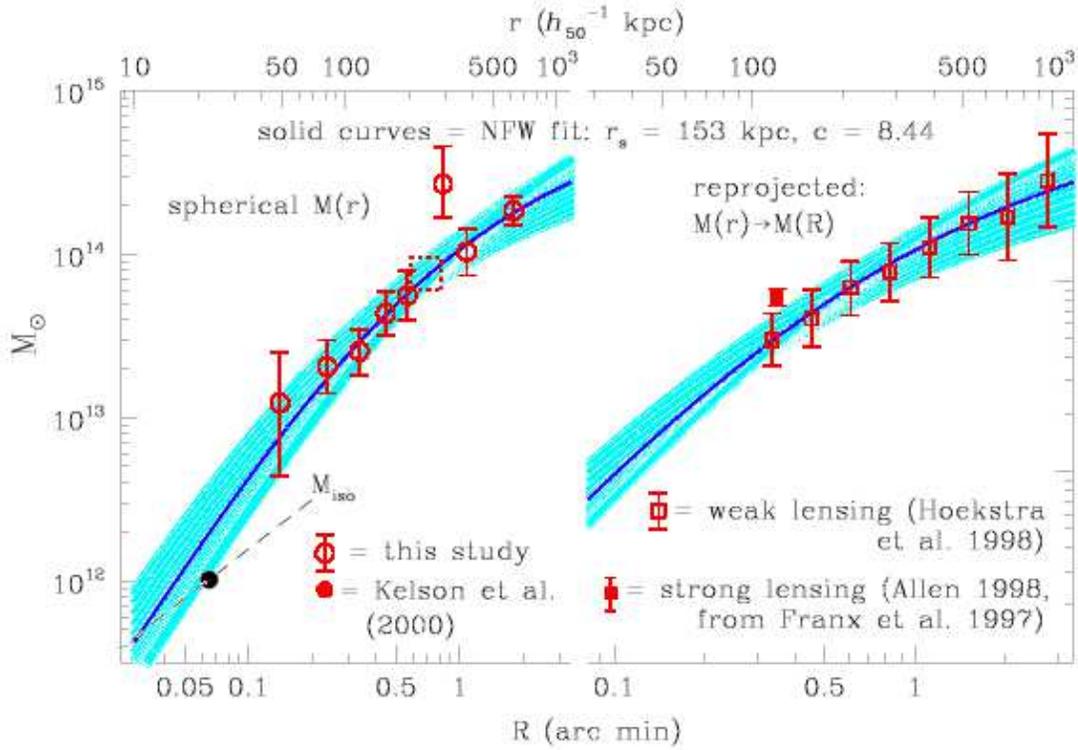,width=15cm}}
\caption{LEFT: Inferred enclosed mass as a function of radius (open circles) 
and best-fit Navarro, Frenk and White (1997; NFW) model (solid line) from Chandra
data on EMSS1358+6245.  RIGHT: Projected mass as a function of projected radius
for the best-fit NFW model obtained from the Chandra data (solid curve) compared
with weak (open squares) and strong (filled squares) lensing results.}
\label{mbautz-B3-fig:fig2}
\end{figure*}

We find that the hot component is isothermal within measurement errors.
The weighted mean temperature is $7.2 \pm 0.1$ keV, in good agreement with
ASCA results.
%
% old version, no non-physical box
%
%   We have therefore assumed a uniform temperature in deriving
%   the mass profile shown in  the left panel of
%   % Figure~\ref{mbautz-B3-fig:fig2}.
%   Figure~3.
%
% new version, w/ non-physical box
%
The resulting mass profile is shown in the left panel of
 Figure~\ref{mbautz-B3-fig:fig2}
(One of the mass profile points, shown as a dashed box, is unphysical
due to stochastic fluctuations in the fitted temperature profile.)
We have fit the so-called universal
density profile of Navarro, Frenk and White (1997; NFW), 
$\frac{\rho(r)}{\rho_0} = x^{-1}(1 +x)^{-2}$ with $x = r/r_s$
(in integral form) to the data; the result is the solid curve in the
left panel of
 Figure~\ref{mbautz-B3-fig:fig2}.
Best-fit values for the model parameters are also shown in
 Figure~\ref{mbautz-B3-fig:fig2}.
The concentration parameter \mbox{$c \equiv r_{vir}/r_{s}$}, where $r_{vir}$ is
the virial radius; see NFW for details.

We have projected our best-fit NFW model mass profile onto the plane of the
sky and, in the right panel of
% Figure~\ref{mbautz-B3-fig:fig2},
Figure~3,
compare it to the weak lensing results of Hoekstra et~al. (1998;
open squares) and the strong lensing result derived by \cite*{mbautz-B3:all98} 
from the
arc discovered by  Franx et al.\ (1997; filled square). The X-ray mass profile
is in excellent  agreement with the weak lensing results.  The strong lensing
point exceeds the both the X-ray and weak lensing mass estimates by a factor
of about 1.6.  We note that the image reconstruction of \cite*{mbautz-B3:fra97}
suggests that an elliptical mass distribution is required to explain the strong
lensing data. The spherical mass estimate dervied by \cite*{mbautz-B3:all98} 
may thus be an overestimate.

\section{Discussion}
\label{mbautz-B3-sec:dis}
The agreement between the X-ray and weak lensing data is strong evidence for
the reliability of the Chandra mass profile. Note that the X-ray data probe 
much smaller radii ($r < 50 $h$_{50}^{-1}$ kpc) than the lensing data, and
so measure the mass distribution at higher densities. They thus may 
constrain dark matter's cross-section for non-gravitational self-interaction. 
In particular, we note that the mean density within the central 
$50$ h$_{50}^{-1}$ 
kpc is $\sim 0.025$ M$_{\odot}$ pc$^{-3}$, comparable to the central density
in dwarf galaxies (Firmani et al.\ 2001).  The flat density profiles in the
cores of such galaxies have been cited, along with other data, as 
possible evidence for self-interacting dark matter 
(e.g. Spergel \& Steinhardt 2000; Dav\'{e} et al.\ 2001).
Assuming the cross-section for any such interaction is identical in the cores
of galaxies and clusters, the mean-free path for interactions will be similar
since the densities are similar. Given the much larger size of cluster cores,
the optical depth for scattering would actually be higher, and the presumed
effects of self-interactions greater, in clusters than in dwarf galaxies, at
least if the systems are of comparable age. 

Indeed, \cite*{mbautz-B3:yos00} have simulated the formation of cluster-sized halos of
self-interacting dark matter with various interaction cross-sections. They find that
even if the dark matter interaction cross-section is as small as 0.1 cm$^{2}$ g$^{-1}$,
i.e., about 50 times smaller than that required to produce flat cores in the galaxy
simulations of \cite*{mbautz-B3:dav01}, a cluster-sized halo would show a flattened core with radius  of order 80 h$_{50}^{-1}$ kpc. 

We find no evidence for a flat core in the density profile
of EMSS1358+6245. A fit of a softened isothermal sphere density profile, with
$ \rho(r) = \frac{ \sigma^{2}}{2 \pi G(r^{2} + r^{2}_{c})} $, where $\sigma$
is  the velocity dispersion and $r_{c}$ the core radius, yields 
$r_{c} = 12^{+30}_{-12}$ h$^{-1}_{50}$ kpc. In this sense, our observations
are inconsistent with any non-gravitational dark matter interaction with a cross-section 
exceeding 0.1 cm$^{2}$ g$^{-1}$. Other Chandra mass profiles published to date,
e.g., for Hydra-A (\cite{mbautz-B3:david01}), Abell 2390 (\cite{mbautz-B3:aef01}), 
Abell 1835 (\cite{mbautz-B3:sch01}) and RXJ1347.5-1145 (\cite{mbautz-B3:asf01}),
all show mass core radii no greater than $\sim 75$ h$_{50}^{-1}$ kpc, and are thus
consistent with our limit on any dark-matter interaction cross-section.

\section{Summary}
\label{mbautz-B3-sec:sum}
A Chandra observation of EMSS1358+6245 has produced a mass profile in good 
agreement with weak lensing results. Both X-ray and weak-lensing mass 
estimates are somewhat lower than the strong lensing estimate at $r \sim 100$ 
h$_{50}^{-1}$ kpc. From the X-ray-derived mass profile we place an upper 
limit of $r_{c} < 42$ h$_{50}^{-1}$  kpc on the radius of any flat core in the
dark matter density profile. Taken together with the self-interacting
dark matter simulations of \cite*{mbautz-B3:yos00}, this result rules out 
a velocity-independent dark matter interaction cross-section large enough
to explain the flat mass density profiles in dwarf galaxies.

\begin{acknowledgements}
This work was supported by NASA under contract NAS8-37716 and NAS8-38252.
\end{acknowledgements}

\end{document}